\documentclass{article}
\usepackage{spconf,amsmath,graphicx}

\usepackage{cite}
\usepackage{amsmath,amssymb,amsfonts}
\usepackage{algorithmic}
\usepackage{graphicx, bm}
\usepackage{textcomp}
\usepackage{xcolor}
\usepackage[]{color-edits}

\usepackage{enumitem}
\setlist{nosep, leftmargin=14pt}

\usepackage{mwe} 

\title{Non-Cartesian Self-Supervised Physics-Driven Deep Learning Reconstruction for Highly-Accelerated Multi-Echo Spiral fMRI}
\name{Hongyi Gu$^{1,2}$,
        Chi Zhang$^{1,2}$, Zidan Yu$^{3}$, Christoph Rettenmeier$^{3}$, V. Andrew Stenger$^{3}$, Mehmet Ak\c{c}akaya$^{1,2}$
        }
\address{$^{1}$Department of Electrical and Computer Engineering, University of Minnesota, Minneapolis, MN, USA\\$^{2}$Center for Magnetic Resonance Research, University of Minnesota, Minneapolis, MN, USA\\$^{3}$Department of Medicine, University of Hawaii, Honolulu, HI, USA. }

\begin{document}
%
\maketitle
\begin{abstract}
Functional MRI (fMRI) is an important tool for non-invasive studies of brain function. Over the past decade, multi-echo fMRI methods that sample multiple echo times has become popular with potential to improve quantification. While these acquisitions are typically performed with Cartesian trajectories, non-Cartesian trajectories, in particular spiral acquisitions, hold promise for denser sampling of echo times. However, such acquisitions require very high acceleration rates for sufficient spatiotemporal resolutions. In this work, we propose to use a physics-driven deep learning (PD-DL) reconstruction to accelerate multi-echo spiral fMRI by 10-fold. We modify a self-supervised learning algorithm for optimized training with non-Cartesian trajectories and use it to train the PD-DL network. Results show that the proposed self-supervised PD-DL reconstruction achieves high spatio-temporal resolution with meaningful BOLD analysis.
\end{abstract}
\begin{keywords}
multi-echo fMRI, non-Cartesian MRI, self-supervised learning, fast MRI
\end{keywords}

\section{Introduction}
\label{sec:intro}
Functional magnetic resonance imaging (fMRI) is a critical tool to non-invasively image and study brain function \cite{bandettini2009s}, and has been integrated to projects such as the Human Connectome Project (HCP) \cite{uugurbil2013pushing}. While conventionally fMRI acquisitions sample a single imaging data per radiofrequency excitation at a pre-specified echo time (TE), over the past decade, there has been substantial interest in multi-echo fMRI acquisitions that sample multiple TEs per excitation \cite{posse2012multi}. These multiple echoes can be potentially used to separate out blood oxygen level dependent (BOLD) signal from other confounding factors, including cardiac or respiratory effects, which are not TE dependent; and may ultimately enable quantification of $T_2^*$ directly \cite{kundu2012differentiating}.

\looseness=-1
Multi-echo fMRI data are typically acquired with Cartesian sampling \cite{posse2012multi}. However, non-Cartesian imaging, especially spiral trajectories, hold promise for a denser sampling of TEs with their efficient coverage of k-space \cite{spiralFMRI}. Yet, for sufficient spatio-temporal resolution, these trajectories need to be acquired without segmentation, requiring very high acceleration rates. At these high rates, conventional reconstruction methods, e.g. parallel imaging as used in HCP \cite{uugurbil2013pushing}. fail to provide reliable reconstructions. Recently, physics-driven deep learning (PD-DL) has demonstrated promise to reconstruct highly-accelerated datasets in the broader MRI community \cite{Hammernik, Hemant, Hosseini_JSTSP}. However, in the context of fMRI, their use has been limited, especially due to a lack of fully-sampled datasets for training at high spatio-temporal resolutions.

In this study, we adapted a self-supervised PD-DL technique \cite{yaman_SSDU_MRM, Yaman_multi_SSDU} to non-Cartesian trajectories and showed its utility in 10-fold accelerated multi-echo spiral fMRI. We performed the self-supervised masking loss in image space using the Toeplitz decomposition for non-Cartesian operators \cite{Ramani, Rapid_Fessler} for fast implementation and efficient training. Results show substantial improvements in image quality over conventional techniques using the proposed approach, while having meaningful activation patterns in subsequent BOLD analysis.

\begin{figure*}[!t]
 \begin{center}
          \includegraphics[trim={0 0 0 0},clip, width=6 in]{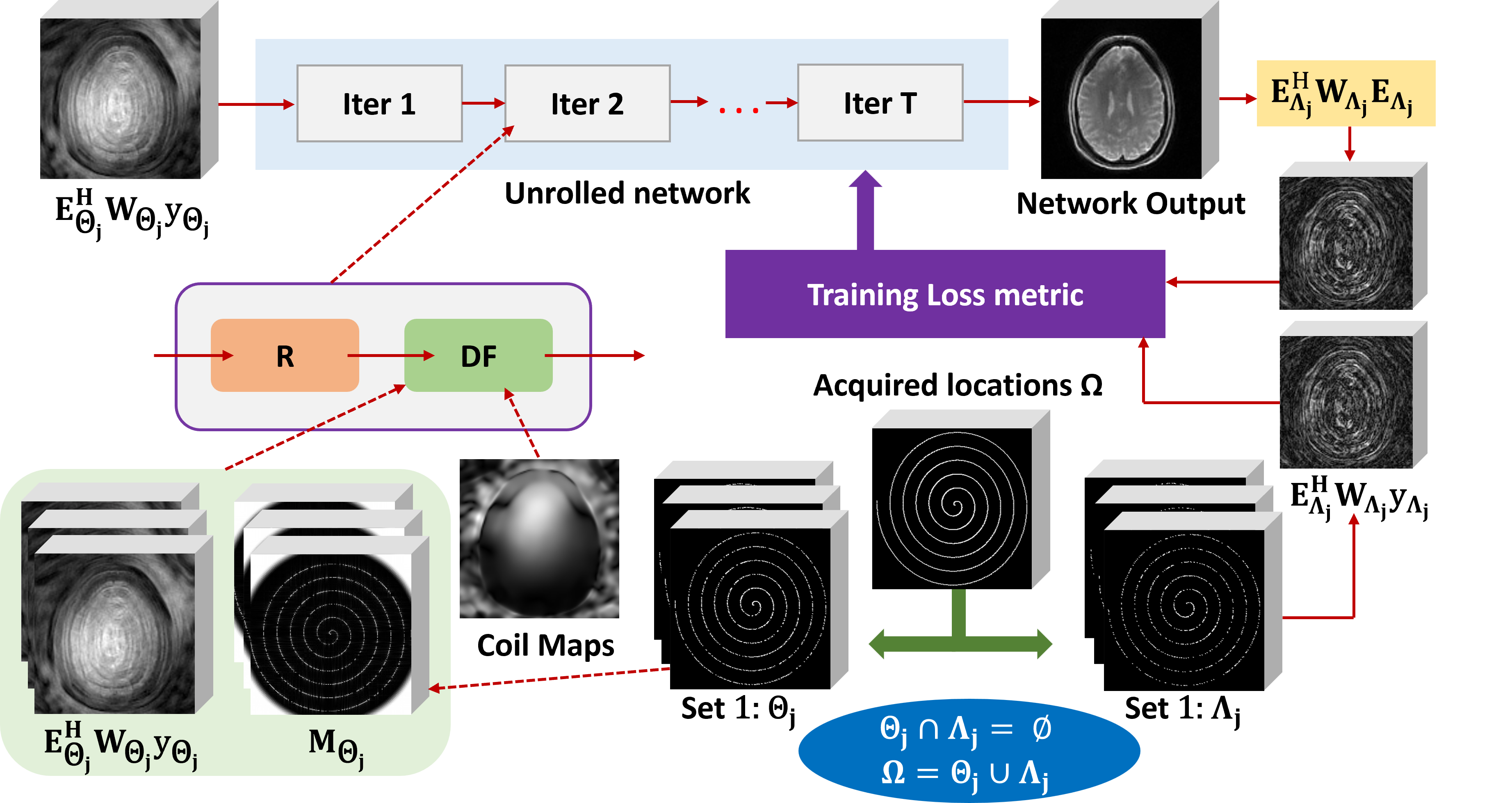}
     \end{center}
      \vspace{-.15cm}
  	\caption{Proposed non-Cartesian implementation for SSDU. The acquired multi-echo non-Cartesian k-space data locations ${\Omega}$ for are split into multiple pairs of disjoint ${\Theta}_j$ and ${\Lambda}_j$, where  ${\Theta}_j \cup {\Lambda}_j={\Omega}$ for all $j$. The unrolled network with $T$ unrolls takes the zero-filled (gridding) solution $\mathbf{E}_{\Theta_j}^H \mathbf{W}_{{\Theta}_j} \mathbf{y}_{{\Theta}_j}$ as input, where $\mathbf{W}_{{\Theta}_j}$ is density compensation based on locations ${\Theta}_j$. Each unrolled block consists of a proximal operator for the regularizer (R) and data fidelity (DF), where the latter uses 
   coil maps and Toeplitz fast operators $\mathbf{M}_{{\Theta}_{j}}$ pre-calculated based on locations ${\Theta}_{j}$. The network output is mapped to the $\Lambda_j$ locations via $\mathbf{E}_{{\Lambda}_{j}}^H \mathbf{W}_{{\Lambda}_{j}} \mathbf{E}_{{\Lambda}_{j}}$ and compared with the gridded image $\mathbf{E}_{{\Lambda}_{j}}^H \mathbf{W}_{{\Lambda}_{j}} \mathbf{y}_{{\Lambda}_{j}}$ on locations ${\Lambda}_{j}$ for loss calculation. The network parameters ${\bm \theta}$ are then updated based on the training loss.
   }
   \label{fig1}
  	\vspace{-.15cm}
\end{figure*}

\section{Materials and methods}

\subsection{Self-Supervised Physics-Driven Deep Learning for Cartesian Datasets}

We first review PD-DL in the Cartesian setup. The inverse problem for Cartesian MRI is given as:
\begin{equation}
\arg \min _{\mathbf{x}}\|\mathbf{E}_{\Omega} \mathbf{x}-\mathbf{y}_{{\Omega}}\|_2^2+\mathcal{R}(\mathbf{x})
\end{equation}
where $\mathbf{x}$ is the image to be reconstructed, $\mathbf{y}_{{{\Omega}}}$ is the acquired Cartesian k-space samples, $\mathbf{E}_{{\Omega}}$ is the forward operator that incorporates coil sensitivities and partial Fourier sampling at set ${\Omega}$, and $\mathcal{R}(\cdot)$ is a regularizer. In PD-DL, a conventional optimization problem for solving this regularized least squares problem \cite{fessler_SPM}, e.g. variable splitting, is unrolled for a fixed number of iterations and solved by alternating between data fidelity (DF) and a regularizer \cite{LeCun, Knoll_SPM}. Here, the proximal operator corresponding to the regularizer is implicitly solved via neural networks, while the DF unit is solved via linear methods such as gradient descent or CG \cite{Hemant}, where relevant DF parameters are learned together with the CNN.

While early works in PD-DL used supervised learning \cite{Hammernik,Hemant}, unsupervised learning methods have also been proposed \cite{akccakaya2022unsupervised}. Among these, multi-mask SSDU \cite{Yaman_multi_SSDU} enables PD-DL training by splitting acquired k-space locations $\Omega$ into multiple pairs of disjoint sets $\{{\Theta}_j, {\Lambda}_j\}$, with ${\Theta}_j \cup {\Lambda}_j={\Omega}$ for all $j$. The samples in ${\Theta}_j$ are used in the DF units of the unrolled network, denoted by $f_{\boldsymbol{\theta}}(\cdot)$ during training, while those in ${\Lambda}_j$, unseen by the network, are used to define a loss. In the Cartesian case, the loss is given by
\begin{equation}
\min _{\boldsymbol{\theta}} \: \mathbb{E} \bigg[ \sum_j \mathcal{L}\left(\mathbf{E}_{{\Lambda}_j} f_{\boldsymbol{\theta}}\left(\mathbf{z}_{{\Theta}_j}\right), \mathbf{y}_{{\Lambda}_j}\right)\bigg] \label{eq:cartSSDU}
\end{equation}
where $\mathcal{L}(\cdot,\cdot)$ is the loss function between the reconstructed k-space and label, and
$\mathbf{z}_{{\Theta}_j} = \mathbf{E}_{{\Theta}_j}^H \mathbf{y}_{{\Theta}_j}$ is the zero-filled solution with acquisition in set ${\Theta}_j$.

\subsection{Proposed Self-Supervised Physics-Driven Deep Learning for Non-Cartesian MRI}
The inverse problem for non-Cartesian MRI involves a generalization to incorporate density compensation \cite{Rapid_Fessler}: 
\begin{equation}
\arg \min _{\mathbf{x}}\left\|\sqrt{\mathbf{W}_{{\Omega}}}\left(\mathbf{E}_{{\Omega}} \mathbf{x}-\mathbf{y}_{{{\Omega}}}\right)\right\|_2^2+\mathcal{R}(\mathbf{x})
\end{equation}
where
$\mathbf{W}_{\Omega}$ refers to density compensation \cite{DCF} of the k-space samples in set $\Omega$, and the rest of the notation is as before. Note the forward operator, $ \mathbf{E}_{{\Omega}}$ utilizes a non-uniform Fourier transform, often referred to as NUFFT \cite{NUFFT}, in addition to the coil sensitivities as aforementioned.

Using variable splitting with quadratic penalty, this problem can be broken into two sub-problems \cite{fessler_SPM}, one involving the proximal operation and the other for DF:
\begin{equation}
\begin{aligned}
\mathbf{z}^{(i)} & =\arg \min _{\mathbf{z}} \mu\left\|\mathbf{x}^{(i-1)}-\mathbf{z}\right\|_2^2+\mathcal{R}(\mathbf{z}) \\
\mathbf{x}^{(i)} & =\arg \min _{\mathbf{x}}\|\sqrt{\mathbf{W}_{{\Omega}}}\left(\mathbf{E}_{{\Omega}} \mathbf{x}-\mathbf{y}_{{{\Omega}}}\right)\|_2^2+\mu\left\|\mathbf{x}-\mathbf{z}^{(i)}\right\|_2^2 \\
& =\left(\mathbf{E}_{{\Omega}}^H \mathbf{W}_{{\Omega}}\mathbf{E}_{{\Omega}}+\mu \mathbf{I}\right)^{-1}\left(\mathbf{E}_{{\Omega}}^H \mathbf{W}_{{\Omega}} \mathbf{y}_{{{\Omega}}}+\mu \mathbf{z}^{(i)}\right)
\end{aligned}
\label{VSQP}
\end{equation}
As before, the first proximal operator sub-problem is solved implicitly with a learnable neural network, and the $\mu$ parameter in DF is also trainable. 

For self-supervised training of such network, we redesign multi-mask SSDU (Fig. \ref{fig1}) to improve its applicability to non-Cartesian MRI reconstruction (referred to as non-Cartesian SSDU).
The main challenge here is related to the time and memory consuming of the NUFFT operation $\mathbf{E}_{\boldsymbol{\Theta}_j}$ and its adjoint $\mathbf{E}_{\boldsymbol{\Theta}_j}^H$ for gridding and regridding processes. Thus, calculation of the loss in k-space as in (\ref{eq:cartSSDU}) leads to prohibitive training times. To this end, we define the following loss function for non-Cartesian self-supervised learning:
\begin{equation}
\min _{\boldsymbol{\theta}} \: \mathbb{E} \Bigg[ \sum_j \mathcal{L}\left(\mathbf{E}_{{\Lambda}_j}^H \mathbf{W}_{{\Lambda}_j} \mathbf{E}_{{\Lambda}_j} f_{\boldsymbol{\theta}}\left(\mathbf{z}_{{\Theta}_j}\right), \mathbf{E}_{{\Lambda}_j}^H \mathbf{W}_{{\Lambda}_j} \mathbf{y}_{{\Lambda}_j}\right) \Bigg],
\end{equation}
where $\mathbf{z}_{{\Theta}_j} = \mathbf{E}_{{\Theta}_j}^H \mathbf{W}_{{\Theta}_j} \mathbf{y}_{{\Theta}_j}$ is zero-filled gridded image with acquisition in set ${\Theta}_j$. Note the major difference to (\ref{eq:cartSSDU}) is that while the loss is still calculated only using the unseen points in $\Lambda_j$, this loss is now calculated on zerofilled gridded image domain data instead of the original k-space to avoid the aforementioned challenges.

Specifically, for this formulation, we recall the Toeplitz decomposition for fast implementation of $\mathbf{E}_{{\Theta}_j}^H \mathbf{W}_{{\Theta}_j} \mathbf{E}_{{\Theta}_j}$ as \cite{Ramani, Rapid_Fessler}:
\begin{equation}
\mathbf{E}_{{\Theta}_j}^H \mathbf{W}_{{\Theta}_j} \mathbf{E}_{{\Theta}_j} \mathbf{x}=\mathbf{S}^H \mathbf{Z}^H \mathcal{F}^H \operatorname{diag}(\mathbf{M}_{{\Theta}_j}) \mathcal{F}\mathbf{Z S} \mathbf{x}
\end{equation}
where $\mathbf{Z}$ is zero-padding of the image matrix $\mathbf{x}$ into double its original size in each dimension, and $\mathbf{Z}^{H}$ is cropping of the image matrix back to its original size; $\mathcal{F}$ is Cartesian Fourier transform operating on the padded image; and the fast operator $\mathbf{M}_{{\Theta}_j}$ is defined as \cite{Rapid_Fessler}
\begin{equation}
\mathbf{M}_{{\Theta}_j}=\mathcal{F} {\mathbf{F}_{{\Theta}_j}^H}\mathbf{W}_{{\Theta}_j} {\mathbf{F}_{{\Theta}_j}} \boldsymbol{\delta}
\end{equation}
where $\boldsymbol{\delta}$ is an impulse image, and $\mathbf{F}_{{\Theta}_j}$ is NUFFT for samples in ${\Theta}_j$ operating on a double-sized image. Note this decomposition has been shown to substantially speed-up training of PD-DL networks with a trade-off in memory \cite{Kooshball}, and is employed for efficient training in this study.

\begin{figure}[t]
\includegraphics[width=\columnwidth]{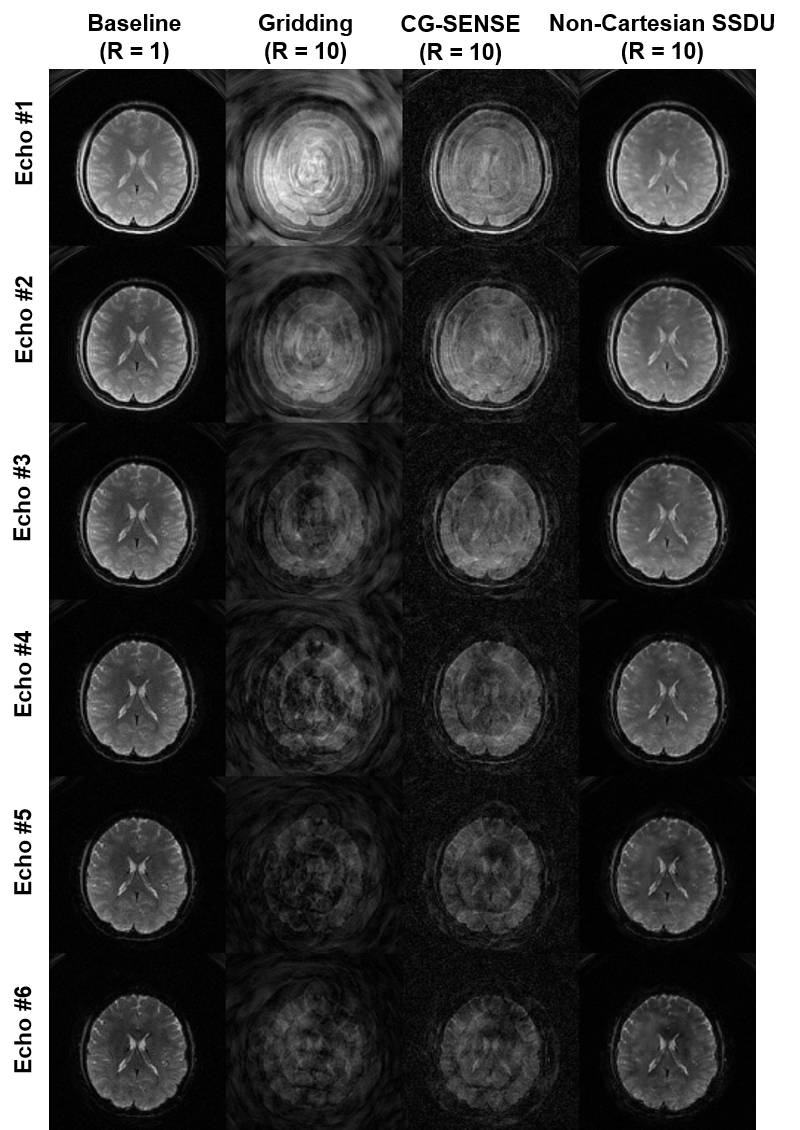}
\caption{Reconstructions from a representative slice of R$=10$ multi-echo spiral data. Gridding and CG-SENSE fail at this high-acceleration rate, while non-Cartesian SSDU gives good reconstruction close to the baseline with 10 spiral arms (R$ = 1$). Note that this dataset is from training acquisitions (described in Section \ref{sec:data}), and not from an fMRI time series, as R$ = 1$ acquisition would be impossible in the latter setup with this spatiotemporal resolution.
   } \label{fig:recon}
\end{figure}

\begin{figure}[!t]
\includegraphics[width=.9\columnwidth]{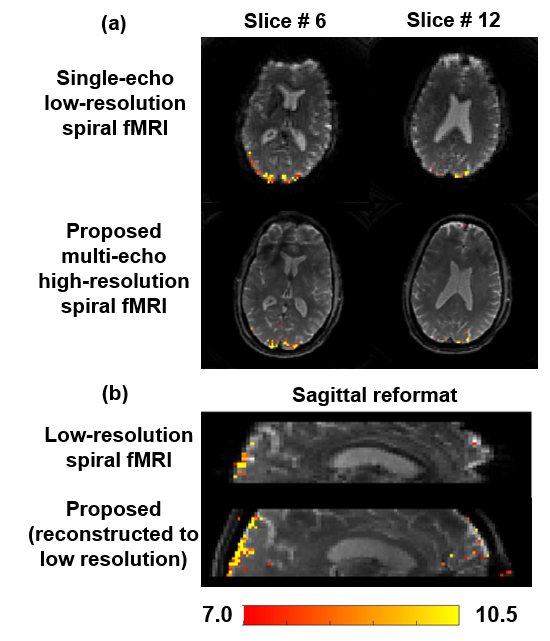}
\caption{Comparison of BOLD sensitivity of single-echo low-resolution spiral fMRI and R$=10$ multi-echo high-resolution spiral fMRI reconstructed with non-Cartesian SSDU. (a) Axial slices show well-preserved activation patterns with the proposed approach, albeit having a 4.7-fold lower baseline SNR compared to the low-resolution acquisition. (b) Sagittal reformats, where the proposed method was reconstructed to the lower resolution, show reliable activation patterns and extent with respect to the baseline low-resolution acquisitions.
} \label{fig:fmri}
\end{figure}

\subsection{Imaging Experiments} \label{sec:data}

All imaging was performed on a 3T system (Prisma; Siemens Healthineers, Germany) with
the relevant institutional review board approvals. The study comprised two sets of data acquisitions. 
The first set was for training data collection on 5 healthy subjects with parameters: flip angle = 20°, TR = 4464 ms, TEs = [3.35, 9.49, 15.63, 21.77, 27.91, 34.05] ms, whole-brain coverage, in-plane FOV = 240 × 240 mm$^2$, resolution = 2 × 2 × 2 mm$^3$. These data were acquired with a single spiral arm (R$=10$ undersampling). As a baseline quality, a corresponding fully-sampled dataset was acquired using 10 spiral arms (R$=1$). Note none of these data included a time-series.

Subsequently, an fMRI experiment was performed on a distinct subject with 10-fold prospective undesampling using the 2D multi-echo sequence with the previous parameters, except partial brain coverage (24 slices) with a 1488ms volume acquisition time. For comparison, a standard single-echo spiral sequence was also acquired with flip angle = 20°, TE/TR = 27.91/1920 ms, in-plane FOV = 240 × 240 mm$^2$, and lower resolution = 3 × 3 × 2 mm$^3$. The fMRI tasks had a total duration of 4:20 minutes, comprising 6 blocks of 20s flickering checkerboard for visual stimulation and separated by 20s resting periods. A general linear model with a canonical hemodynamic response function was used for BOLD analyses, where first and second order temporal trends in the data were removed. Multi-echo data was prepared using weighted echo summation \cite{BOLD} based on local $T_2^*$.

\subsection{PD-DL Implementation Details}
For PD-DL, variable splitting with quadratic penalty was unrolled for $T = 10$ iterations, with DF unit solved via 15 conjugate gradient iterations \cite{Hemant}. DF was solved for each echo of the slice individually, while the regularizer acted on all echo images jointly.
Coil maps were estimated from central k-space region of the first echo of fully-sampled baseline data, which was also performed for calibration purposes prior to the fMRI experiments. 

The non-Cartesian SSDU training used 7 masks applied to the central 60 slices from 4 subjects with 10-fold accelerated single-shot acquisition (without the time series). Each mask was generated via uniform random selection along the singles-shot spiral arm, with center 32 samples used in each ${\Theta}_j$. For each mask we use 60\%-40\% separation for ${\Theta}$-${\Lambda}$ \cite{Yaman_multi_SSDU}. Training was performed with learning rate of $5 \cdot 10^{-4}$ 
using a mixed $\ell_1$-$\ell_2$ loss \cite{Knoll_SPM}. Testing was done on all acquired ${\Omega}$, corresponding to a 10-fold accelerated single-shot acquisition of a different subject from training without time series (in the first set of data) or a 10-fold accelerated in-vivo fMRI acquisition (in the second set of data). 

\section{Results}
Figure \ref{fig:recon} shows representative reconstruction results of a 10-fold accelerated multi-echo spiral sequence. As aforementioned, these data come from the first set of acquisitions, which do not contain a time series or an fMRI experiment. As such, it is possible to acquire a baseline image (R$=1$) with a segmented acquisition containing 10 spiral arms. At the target R$=10$ acceleration, gridding and CG-SENSE yield poor reconstruction quality as expected. PD-DL trained with non-Cartesian SSDU provides substantial improvements, closely matching the R$=1$ baseline.

Figure \ref{fig:fmri} shows comparison of BOLD sensitivity analysis, overlaid on images from single-echo low-resolution spiral fMRI and the proposed R$=10$ multi-echo high-resolution spiral fMRI reconstructed using non-Cartesian SSDU. Figure \ref{fig:fmri}a shows comparisons in axial slices. Note that multi-echo high-resolution spiral fMRI has improved in-plane resolution (2 × 2 mm$^2$) than single-echo low resolution spiral fMRI (3 × 3 mm$^2$), and the two acquisitions also have different TE and TR. Thus they have overall different contrast, while the single-echo low resolution spiral fMRI has about 4.7-fold higher baseline SNR. Nonetheless, BOLD signal is well-preserved in the proposed method using non-Cartesian SSDU, albeit with diminished sensitivity. Figure \ref{fig:fmri}b shows comparisons in sagittal view, where the proposed method was reformatted to match the resolution of the low-resolution spiral fMRI. In this fairer comparison, activation patterns and extent of activation are more closely matched between the two acquisitions, highlighting the potential of the proposed PD-DL reconstructed multi-echo fMRI approach. 

\section{Discussion and conclusion}
In this work, we developed a non-Cartesian version of SSDU for efficient training of non-Cartesian PD-DL reconstructions, and showed its potential in highly-accelerated multi-echo spiral fMRI. The results indicate that 10-fold accelerated multi-echo spiral fMRI may be feasible using PD-DL reconstruction, where BOLD analysis sensitivity was largely preserved with respect to low-resolution single-echo spiral fMRI acquisitions, albeit the notably higher SNR for the latter acquisition. 

While this work establishes the feasibility of multi-echo spiral fMRI, further sequence changes are needed for higher acceleration in the slice direction to enable whole-brain coverage, via simultaneous multi-slice imaging or 3D acquisitions \cite{uugurbil2013pushing}. The latter approach may necessitate advances in large-scale processing as outlined in \cite{Kooshball}, and will be investigated in future studies. Finally, we note that the fMRI acquisitions in this study utilized a different TR than the training data, and the performance of PD-DL may further be improved by training on matched datasets \cite{hammernik2023physics}.

\section{Acknowledgments}
This work was partially supported by NIH R01EB032830, NIH P41EB027061, NIH P20GM139753, NIH R01EB028627. The first three authors contributed equally.
\label{sec:acknowledgments}

\section{COMPLIANCE WITH ETHICAL STANDARDS}
This research study was performed in line with the principles of the Declaration of Helsinki. Approval was granted by the Institutional Review Board of the University of Hawaii.
\label{sec:ethics}

\bibliographystyle{IEEEbib}
\bibliography{reference}

\end{document}